\documentclass[12pt]{iopart}
\usepackage{iopams}
\newtheorem{defi}{Definition}
\newtheorem{lem}[defi]{Lemma}
\newtheorem{thm}[defi]{Theorem}

\def\QED{\mbox{\rule[0pt]{1.5ex}{1.5ex}}}

\def\endproof{\hspace*{\fill}~\QED\par\endtrivlist\unskip}
\def\rank{\mathop{\rm rank}\nolimits}

\def\Tr{\mathop{\rm Tr}\nolimits}

\def\SL{\mathop{\rm SL}\nolimits}

\def\Id{I}

\def\real{\mathbb{R}}
\def\complex{\mathbb{C}}

\def\Label#1{\label{#1}\ [\ #1\ ]\ }
\def\Label{\label}
\begin{document}
\jl{1}
\title[Optimal sequence of quantum measurements]{Optimal 
sequence of quantum measurements\\
in the sense of Stein's lemma\\
in quantum hypothesis testing}
\author{Masahito Hayashi\dag}
\address{\dag\
Laboratory for Mathematical Neuroscience, 
Brain Science Institute, RIKEN\\
2-1, Hirosawa, Wako, Saitama 351-0198, Japan \\
e-mail masahito@brain.riken.go.jp}
\begin{abstract}
We derive a necessary and sufficient condition
for a sequence of quantum measurements to achieve the optimal performance 
in quantum hypothesis testing.
We discuss what quantum measurement we should perform in 
order to attain the optimal exponent of the second error
probability under the condition that the first error 
probability goes to $0$.
As an asymptotically optimal measurement,
we propose a projection measurement characterized 
by the irreducible representation theory of the 
special linear group $\SL( {\cal H})$.
Specially, in spin 1/2 system,
it is realized by the simultaneous measurement
of the total momentum and a momentum of a specified direction.
As a byproduct,
we obtain another proof of quantum Stein's lemma.
In addition, an asymptotically optimal meassurement is constructed in the 
quantum Gaussian case,
and it is physically meaningful.
\end{abstract}
\submitted
\maketitle

\section{Introduction}\Label{s1}
Deciding the true quantum state based on the two hypotheses
is called quantum hypothesis testing,
which is one of the most fundamental problems
among quantum information theory because of the following reasons.
The difficulty derived from non-commutativity
of matrices (operators) appears as a simple form in this problem.
This problem can be applied to 
other related topics in quantum information,
for example, quantum channel coding \cite{OD,Nacon,HN,ON1},
distillable entanglement \cite{Ra},
quantum estimation \cite{Haya3},
quantum universal variable-length source coding \cite{HM},
and quantum coin tossing \cite{MSK}.

When the null hypothesis is the tensor product of a certain 
quantum state $\rho$ and the alternative hypothesis is that of
another quantum state $\sigma$,
we sometimes focus on the asymptotic behaviors of the
first error probability
(we reject the null hypothesis though it is correct)
and the second error probability
(we accept the null hypothesis though it is incorrect).
Hiai and Petz \cite{HP}, and Ogawa and Nagaoka \cite{ON}
discussed the optimal second error exponent
under the assumption that the first error 
probability is less than a certain constant $\epsilon \,> 0$.
Combining their results, we obtain that
the optimal second error exponent is independent of $\epsilon \,> 0$,
and coincides with the quantum relative entropy.
Hiai and Petz \cite{HP} proved the direct part, i.e.,
the attainability of the quantum relative entropy,
and Ogawa and Nagaoka \cite{ON} proved the converse part i.e.,
the impossibility for surpassing the quantum relative entropy.
The converse part was simplified by Nagaoka \cite{Nacon}.
In addition, 
the quantum relative entropy
coincides with the optimal second error exponent 
under the condition that the first error probability 
asymptotically goes to $0$.
Moreover, Ogawa and Hayashi \cite{OH} discussed the second error 
exponent under the
constant constraint for the first error exponent.

We divide this testing process into two parts:
One is the quantum part i.e., the quantum measurement process.
When $\rho$ and $\sigma$ are non-commutative,
the choice of this quantum measurement is difficult and essential.
The other is the classical part i.e.,
the classical data processing.
In this paper, we focus on the former process i.e.,
we study what kind of measurement 
is suitable in order to achieve the optimal second error exponent.
As mentioned in section \ref{s6},
it is sufficient for this kind hypothesis testing,
to discuss our quantum measurement of a certain class.
We derive a necessary and sufficient condition 
for a quantum measurement to attain the optimal second error exponent
among this class.
This condition depends on the alternative hypothesis
$\sigma$, and is almost independent of the null one $\rho$.
As a byproduct, we obtain another proof of quantum Stein's lemma.

In our setting, the unknown state is a tensor product state,
but our measurement is not necessarily tensor product.
Therefore, in order to treat the classical part
we need to discuss our data-processing after 
our measurement as a classical hypothesis testing with two
general sources.
In classical information theory, by using the information-spectrum 
method, Han \cite{Han,Han2} studied hypothesis testing 
based on such a general setting.
We apply it to our proof of the main result, and such an application
to quantum hypothesis testing was 
initiated by Nagaoka \cite{Naga,Naga2}.
However, this paper is organized so that
the reader can understand the statement of 
the main result without any knowledge of the information-spectrum method.
A quantum version of this method was discussed by 
Nagaoka and Hayashi \cite{NH},
but it is not treated in this paper because it
is not directly related to this issue.
This work was motivated by Nagaoka's \cite{Naga,Naga2} earlier works.
Unfortunately, these Nagaoka's papers were written in Japanese,
but Nagaoka and Hayashi \cite{NH} contains a part of results by Nagaoka
\cite{Naga,Naga2}.

This paper is organized as follows.
In section \ref{s3}, we formulate quantum hypothesis testing 
with tensor product states as an asymptotic problem.
In order to discuss our quantum asymptotic setting,
we prepare some non-asymptotic characterizations in section \ref{s4}.
After these preparations, we state the main results
i.e., we characterize 
a quantum measurement to attain the optimal second error exponent
in section \ref{s6}.
We treat quantum Gaussian states as a special example
of the infinite-dimensional case in section \ref{s7},
while we assume that the dimension of the Hilbert space of interest 
is finite in section \ref{s6}.
In the quantum Gaussian case,
we give an asymptotically optimal measurement whose physical 
interpretation is clear.
In order to prove our main theorem,
we have to discuss a general sequence of classical information sources.
Thus, in section \ref{s2}, 
by using the information-spectrum method,
we prepare a lemma which is applicable to such a general sequence,
and then apply it to our issue.
we prove the main theorem in section \ref{sp} with 
a help of this lemma.
We use some fundamental inequalities in section \ref{sp}, and
these inequalities are given in section \ref{snew}.
Moreover we assume a fundamental fact in section \ref{s6},
this fact is proven from a representation viewpoint
in section \ref{s5}.

\section{Asymptotic formulation of quantum hypothesis testing}\Label{s3}
Let ${\cal H}$ be the Hilbert space of
interest, and ${\cal S}({\cal H})$
be the set of density matrices on ${\cal H}$.
When we perform a measurement corresponding to 
POVM (Positive Operator Valued Measure) 
$M= \{ M_i \}$ to a system in the state $\rho$,
the data obeys the probability 
${\rm P}_{\rho}^M=
\{ {\rm P}_{\rho}^M(i) = \Tr M_i \rho\}$.
In particular, the POVM $M= \{M_i\}$ is called a PVM (Projection
Valued Measure) if each $M_i$ is a projection.
In the hypothesis testing, the testing is described by a 2-valued POVM 
$\{ M_a , M_r \}$, where $M_a$ corresponds to acceptance
and $M_r$ corresponds to rejection.
Similarly, an operator $A$ satisfying $0 \le A \le \Id $
is called a {\it test},
and is identified with the POVM $\{ M_a , M_r \}=
\{ A, \Id - A\}$.

Now, we study the quantum hypothesis testing problem
for the null hypothesis $H_0:
\rho^{\otimes n}\in {\cal S}({\cal H}^{\otimes n})$
versus the alternative hypothesis $H_1:
\sigma^{\otimes n}\in {\cal S}({\cal H}^{\otimes n})$,
where $\rho^{\otimes n}$ and $\sigma^{\otimes n}$
are the $n$th-tensor powers of arbitrarily
given density operators
$\rho$ and $\sigma$ in ${\cal S}({\cal H})$.
In the sequel, an operator $A^n$ on ${\cal H}^{\otimes n}$
satisfying $0 \le A^n \le \Id$
or a sequence $\{ A^n \}$ of such operators, is called a {\it test}.
For a test $A^n$ the probabilities of the first and the second are,
respectively, defined by
\begin{eqnarray*}
\alpha_n(A^n)= \Tr \rho^{\otimes n}( \Id - A^n) 
\hbox{ and }
\beta_n(A^n)= \Tr \sigma^{\otimes n} A^n. 
\end{eqnarray*}
We can understand that $\alpha_n(A^n)$
is the probability of erroneously
rejecting $\rho^{\otimes n}$
though $\rho^{\otimes n}$
is true and $\beta_n(A^n)$ is the error probability
of erroneously accepting $\rho^{\otimes n}$
although $\rho^{\otimes n}$ is not true.
We discuss the trade-off of the two type error probabilities
for two $n$-tensor product states

The following is known as quantum Stein's lemma.
\begin{thm}\Label{Stein}
For any $1 \,> \epsilon \,> 0$, the equation 
\begin{eqnarray*}
- \lim_{n \to \infty}
\frac{1}{n} \log \beta_n^*(\epsilon)
=D(  \rho\| \sigma):= \Tr \rho (\log \rho - \log \sigma)
\end{eqnarray*}
holds, where 
$\beta_n^*(\epsilon)$ is the minimum second error probability
under the constraint that the first error probability
is less than $\epsilon$ i.e., 
\begin{eqnarray*}
\beta_n^*(\epsilon)
:= \min \{ \beta_n(A^n)|
0 \le A^n \le \Id , \alpha_n(A^n) \le \epsilon \}.
\end{eqnarray*}
\end{thm}
The part $\ge$ was proven by Hiai and Petz \cite{HP}.
Its infinite-dimensional case was proved by Petz \cite{Petz}.
The part $\le$ was proven by Ogawa and Nagaoka \cite{ON}.

In order to discuss the above theorem,
we define the two following quantities
\begin{eqnarray*}
\fl B ( \rho \| \sigma)
:=&
\sup \left\{
\lambda \left|
\exists \vec{A},
\lim_{n \to \infty} \alpha_n(A^n) = 0,
 \limsup_{n \to \infty}
\frac{1}{n} \log \beta_n(A^n) \le -\lambda 
\right. \right\} , \\
\fl B^{\dagger} ( \rho \| \sigma)
:=&
\sup \left\{
\lambda \left|
\exists \vec{A},
\limsup_{n \to \infty} \alpha_n(A^n) \,< 1,
\limsup_{n \to \infty}
\frac{1}{n} \log \beta_n(A^n) \le -\lambda 
\right.\right\}  
\end{eqnarray*}
i.e.,
the former is the optimal second error exponent under
the constraint that the first error probability goes to $0$
while the later is the optimal second error exponent under
the other constraint that the first error probability does not go to $1$.
Thus, quantum Stein's lemma (Theorem \ref{Stein}) is equivalent to
the following relations
\begin{eqnarray*}
B ( \rho \| \sigma)
=B^{\dagger} ( \rho \| \sigma)
=D ( \rho \| \sigma).
\end{eqnarray*}

In the following, we divide this testing process into the two parts
i.e., the quantum part and the classical part.
The quantum part corresponds to the choice of
a sequence $\vec{M}:=\{ M^n \}$ of POVMs,
and the classical part does to the decision process from 
the classical data.
Now, we fix a sequence $\vec{M}:=\{ M^n \}$ of POVMs,
and denote the data set of $M^n$ by $\Omega_n$.
Then, this classical part is formulated as
the classical hypothesis testing
with the null hypothesis ${\rm P}_{\rho^{\otimes n}}^{M^n}$
and the alternative hypothesis ${\rm P}_{\sigma^{\otimes n}}^{M^n}$.
In this setting, a test is described by a function $T^n$
from $\Omega_n$ to $[0,1]$
instead of an operator $A^n$ satisfying $0 \le A^n \le I$.
In particular, when a test $T^n$ is a test function
with the support $S^n$,
this test is equivalent to the test with the accept region $S^n$.
Similar to $B ( \rho \| \sigma)$ and $B^{\dagger} ( \rho \| \sigma)$,
we define 
\begin{eqnarray*}
\fl B^{\vec{M}} ( \rho \| \sigma)
:=&
\sup \left\{
\lambda \left|
\exists \vec{A},
\lim_{n \to \infty} {\rm E}^{M^n}_{\rho^{\otimes n}} (1- T^n)
 = 0,
 \limsup_{n \to \infty}
\frac{1}{n} \log {\rm E}^{M^n}_{\sigma^{\otimes n}} (T^n)
 \le -\lambda 
\right. \right\} , \\
\fl B^{\dagger,\vec{M}} ( \rho \| \sigma)
:=&
\sup \left\{
\lambda \left|
\exists \vec{A},
\limsup_{n \to \infty} {\rm E}^{M^n}_{\rho^{\otimes n}} (1- T^n)
\,< 1,
\limsup_{n \to \infty}
\frac{1}{n} \log {\rm E}^{M^n}_{\sigma^{\otimes n}} (T^n)
\le -\lambda 
\right.\right\}  ,
\end{eqnarray*}
where ${\rm E}^M_{\rho}$ denotes the expectation regarding the 
probability ${\rm P}^M_{\rho}$.
We can easily check that
\begin{eqnarray*}
B ( \rho \| \sigma) =
\sup_{\vec{M}:\hbox{\small POVMs}} B^{\vec{M}} ( \rho \| \sigma)
\le 
B^{\dagger} ( \rho \| \sigma) =
\sup_{\vec{M}:\hbox{\small POVMs}} B^{\dagger,\vec{M}} ( \rho \| \sigma).
\end{eqnarray*}
Therefore, 
Theorem \ref{Stein} is equivalent to
the following relations
\begin{eqnarray}
\sup_{\vec{M}:\hbox{\small POVMs}} B^{\vec{M}} ( \rho \| \sigma)
=
\sup_{\vec{M}:\hbox{\small POVMs}} B^{\dagger,\vec{M}} ( \rho \| \sigma)
=D ( \rho \| \sigma). \Label{8.3}
\end{eqnarray}
In this paper, 
we focus on a sequence 
$\vec{M}:= \{ M^n \}_{n=1}^{\infty}$
of POVMs that satisfies the condition
\begin{eqnarray}
B^{\vec{M}} ( \rho \| \sigma)
= D ( \rho \| \sigma ) ,\Label{3}
\end{eqnarray}
and call such a sequence $\vec{M}$ of POVMs(PVMs)
an {\it optimal sequence of POVMs(PVMs) in the sense
of Stein's lemma}.
The main issue is a characterization 
of a sequence 
$\vec{M}:= \{ M^n \}_{n=1}^{\infty}$
of POVMs that satisfies the condition (\ref{3}).
As mentioned in section \ref{s6}, our characterization of 
such an optimal sequence is independent of the null hypothesis
$\rho$, and depends only on the alternative hypothesis $\sigma$.
Of course, in section \ref{s6}, 
we construct such an optimal sequence.
Indeed, if a sequence 
$\vec{M}$ of POVMs satisfies the condition (\ref{3}),
there exists 
a sequence $\vec{A}:=\{A^n \}$ of tests satisfying 
\begin{eqnarray}
\lim_{n \to \infty} \alpha_n(A^n) = 0 , \quad
-\lim_{n \to \infty}\frac{1}{n} \log \beta_n(A^n)
= D ( \rho \| \sigma ) - \epsilon , \Label{2}
\end{eqnarray}
for any $\epsilon \,> 0$.
In the following, we assume that the dimension of ${\cal H}$
is finite ($k$) and the inverse $\sigma^{-1}$
of $\sigma$ exists.

\section{Non-asymptotic characterization of PVMs}\Label{s4}
In order to treat the condition (\ref{3}), 
we need some characterizations concerning PVMs 
in the non-asymptotic setting.
One may think that these characterizations
have no relation with the condition (\ref{3}), 
but they are essential for our issue.

A state $\rho$ is called {\it commutative} with 
a PVM $E(=\{ E_i \})$ on ${\cal H}$ 
if $\rho E_i = E_i \rho $ for any index $i$.
The spectral decomposition of any operator $X$
can be regarded as a PVM and it is denoted by $E(X)$.
In particular, we have $E(\sigma)= E(\log \sigma)$.
The map ${\cal E}_E$ 
with respect to a PVM $E$ is defined as:
\begin{eqnarray*}
 {\cal E}_E : \rho \mapsto \sum_{i} E_i \rho E_i , 
\end{eqnarray*}
which is 
a linear map from the set of Hermite operators to itself.
Note that the state ${\cal E}_E(\rho)$ is commutative with a PVM $E$.
The number $\sup_i \rank E_i$ of a PVM $E= \{E_i\}$
is an important quantity in the following,
and is denoted by $w(E)$.
Next, we focus on two PVMs 
$E (=\{ E_i \}_{i \in I}) ,F(=\{ F_j \}_{j \in J})$.
We write $E \le  F$ if 
there exists a subset $(F/E)_i$ of the index set $J$
such that $E_i = \sum_{j \in (F/E)_i} F_j$ for any index $i \in I$.
If a PVM $F=\{ F_j \}$ is commutative with a PVM $E=\{ E_i \}$, then
we can define the PVM $F \times E= \{ F_j E_i \}$,
which satisfies $F \times E \ge E$ and $F \times E \ge F$,
and can be regarded as the simultaneous measurement of $E$ and $F$.

\begin{lem}\Label{LL1}
If $\rho$ and $\sigma$ are commutative with
a PVM $E$, then the equation
\begin{eqnarray*}
\inf\left\{
\beta ( A ) \left |
\alpha (A)  \le \epsilon \right. \right\} \\
\lo = \inf\left\{
\beta ( A ) \left |
\exists M : {\rm PVM }
,\quad M\ge E , 
M \ge E(A)  , 
\alpha (A)  \le \epsilon , 
w(M)=1
\right. \right\} 
\end{eqnarray*}
holds.
\end{lem}
\begin{proof}
For any $A$,
the relations $\beta({\cal E}_E(A))=\beta(A)$ and $
\alpha({\cal E}_E(A))=\alpha(A)$
hold.
Since the PVM $E({\cal E}_E(A))$ commutes
the PVM $E$,
there exists a PVM $M$ such that 
$M \ge E, M \ge E({\cal E}_E(A))$ and $w(M)=1$.
\end{proof}
Indeed, if a test $A$ and a PVM $M$ satisfy $M \ge E(A)$, 
the test $A$ is performed by combining the quantum measurement $M$
and a suitable data processing.
Therefore, when $\rho$ and $\sigma$ are commutative with
a PVM $E$,
we may discuss only PVMs $M$
satisfying $M \ge E$ i.e.,
we can restrict our tests.

\section{Main Result}\Label{s6}
In this section,
we discuss the condition (\ref{3}) under 
the assumption that there exist PVMs $E^n$ such that
each PVM $E^n$ is commutative
with states $\sigma^{\otimes n}$ and $\rho^{\otimes n}$
and $w(E^n) \le (n+1)^{k-1}$.
This existence is proven by the representation theory in section \ref{s5}.
Since,
it follows from Lemma \ref{LL1} that
we may treat only a PVM satisfying $M^n \ge E^n$ and $w(M^n)=1$,
we obtain 
\begin{eqnarray}
\fl B(\rho\|\sigma)
=\sup_{\vec{M}:\hbox{\small POVMs}} B^{\vec{M}}
(\rho\|\sigma)
=\sup_{\vec{M}=\{M^n\}:\hbox{\small PVMs s.t. } M^n \ge E^n,
w(M^n)=1} 
B^{\vec{M}}(\rho\|\sigma) \nonumber
\\
\fl B^{\dagger}(\rho\|\sigma)
=\sup_{\vec{M}:\hbox{\small POVMs}} B^{\dagger,\vec{M}}
(\rho\|\sigma)
=\sup_{\vec{M}=\{M^n\}:\hbox{\small PVMs s.t. } M^n \ge E^n,
w(M^n)=1} 
B^{\dagger,\vec{M}}(\rho\|\sigma),
\Label{e919}
\end{eqnarray}
i.e., we can discuss 
$B^{\vec{M}}(\rho\|\sigma)$ and 
$B^{\dagger,\vec{M}}(\rho\|\sigma)$ only of 
a sequence $\vec{M}$ satisfying
the condition $M^n \ge E^n$, $w(M^n)=1$.

Therefore, our main issue is the asymptotic behavior 
of the variable $\frac{1}{n}\log
\frac{{\rm P}_{\rho^{\otimes n}}^{M^n}}
{
{\rm P}_{\sigma^{\otimes n}}^{M^n}}$ 
for the probability distribution
${\rm P}_{\rho^{\otimes n}}^{M^n}$
under the condition $M^n \ge E^n$, $w(M^n)=1$.
\begin{thm}\Label{thm9}
For any sequence $\vec{M} = \{M^n\}$ of PVMs 
satisfying $M^n \ge E^n$, $w(M^n)=1$,
the relation 
\begin{eqnarray}
B^{\vec{M}}(\rho\|\sigma)
= D ( \rho \| \sigma ) \Label{727}
\end{eqnarray}
holds, if and only if
the variable $- \frac{1}{n}\log {\rm P}_{\sigma^{\otimes n}}^{M^n}$ 
converges to $- \Tr \rho \log \sigma$ in the probability
distribution ${\rm P}_{\rho^{\otimes n}}^{M^n}$.
\end{thm}
For example,
if a PVM $M^n$ is commutative with $\sigma^{\otimes n}$ and satisfies
$M^n \ge E^n $, $w(M^n)=1$,
the equations
\begin{eqnarray}
\fl \sum_{i}{\rm P}_{\rho^{\otimes n}}^{M^n}(i)\left(
\frac{1}{n}\log {\rm P}_{\sigma^{\otimes n}}^{M^n}(i)
- \Tr \rho \log \sigma \right)^2
=
\Tr {\cal E}_{M^n}(\rho^{\otimes n})
\left(
\frac{1}{n}\log {\cal E}_{M^n}(\sigma^{\otimes n})
- \Tr \rho \log \sigma \right)^2 \nonumber \\
\fl =
\Tr \rho^{\otimes n} \left(
\frac{1}{n}
\log \sigma^{\otimes n} -  \Tr \rho \log \sigma \right)^2
=
\Tr \rho^{\otimes n} \left(
\frac{1}{n}
(\log \sigma)^{(n)} -  \Tr \rho \log \sigma \right)^2  \nonumber\\
\fl = 
\frac{1}{n}
\Tr \rho \left(\log \sigma - \Tr \rho \log \sigma  \right)^2
 \Label{dd}
\end{eqnarray}
hold.
The equation (\ref{dd}) implies that the variable
$\frac{1}{n}\log {\rm P}_{\sigma^{\otimes n}}^{M^n}$ converges to 
$\Tr \rho \log \sigma$ in probability.
Therefore, it is optimal in the sense of Stein's lemma.
This PVM coincides with the PVM proposed by Hayashi\cite{Haya}.
In particular, as guaranteed in section \ref{s5},
in spin 1/2 system,
$E^n \times E({\sigma^{\otimes n}})$ 
can be regarded as a simultaneous measurement 
of the total momentum and a momentum of the specified direction.
As a byproduct, we can prove the following theorem.
\begin{thm}\Label{thm1}
Any sequence $\vec{M}$ of POVMs satisfies 
\begin{eqnarray*}
D(\rho\|\sigma) \ge 
B^{\dagger,\vec{M}}(\rho\|\sigma).
\end{eqnarray*}
\end{thm}
Theorems \ref{thm9} and \ref{thm1} yield
relation (\ref{8.3}), 
which is equivalent to quantum Stein's lemma.

\section{Quantum Gaussian states}\Label{s7}
In this section,
we discuss a quantum hypothesis testing 
whose hypotheses are quantum Gaussian states $\rho_\theta$
on an infinite dimensional space $L^2(\real)$:
\begin{eqnarray*}
\rho_\theta:=
\frac{1}{\pi \overline{N}} \int_{\complex}
| \alpha \rangle \langle \alpha |
e^{-\frac{|\alpha-\theta|^2}{\overline{N}}}\,d^2 \alpha
, \quad \forall \theta \in \complex,
\end{eqnarray*}
where 
we define the boson coherent vector $| \alpha \rangle:=
e^{- \frac{|\alpha|^2}{2}}
\sum_{k=0}^{\infty}\frac{\alpha^{k}}{\sqrt{k !}}|k \rangle$
and $|k \rangle$ is the number vector on $L^2(\real)$.
Now, we propose a suitable PVM.
This PVM not only attains the 
optimal exponents of the second error probability,
but also has an intuitive physical meaning.
When the null hypothesis is
the state $\rho_{\theta_0}^{\otimes n}$ and the alternative hypothesis is
the state $\rho_{\theta_1}^{\otimes n}$,
our PVM is constructed as follows.
First, we perform the unitary evolution as
\begin{eqnarray*}
\rho_{\theta}^{\otimes n} \mapsto
\rho_{\theta-\theta_1}^{\otimes n}.
\end{eqnarray*}
Second, we perform the unitary evolutions 
\begin{eqnarray*}
\rho_{\theta-\theta_1}^{\otimes n} \mapsto
\rho_{\sqrt{n}(\theta-\theta_1)} \otimes 
\rho_{0}^{\otimes (n-1)}.
\end{eqnarray*}
The constructions of these unitary evolutions
were mentioned in Appendix F of Hayashi \cite{Haya3}.
Finally, we perform a number detection
$\{ |k \rangle \langle k |\}_{k=0}^\infty$ to the system whose state is
$\rho_{\sqrt{n}(\theta-\theta_1)}$ and denote the 
final data (this PVM) by $k$ ($M^n_{\theta_1}$).
Following to section 7.1 in Hayashi \cite{Haya3} and its appendixes,
we obtain
\begin{eqnarray*}
{\rm P}^{M^n_{\theta_1}}_{\rho_{\theta_0}^{\otimes n}}
\left\{ \left| \sqrt{\frac{k}{n}} - |\theta_0 - \theta_1| \right| \,> \epsilon \right\}
\to 0 \\
\frac{-1}{n} \log {\rm P}^{M^n_{\theta_1}}_{\rho_{\theta_1}^{\otimes n}}
\left\{ \sqrt{\frac{k}{n}}  \ge  |\theta_0 - \theta_1 | \right\}
\to
|\theta_0 - \theta_1 |^2 \log \left(1+ \frac{1}{N}\right)
= D( \rho_{\theta_0} \|\rho_{\theta_1} )
\end{eqnarray*}
for any $\epsilon \,> 0$, and any $\theta \in \complex$.
Therefore, 
when we choose the acceptance region as
$\left\{  \left| \sqrt{\frac{k}{n}} - |\theta_0 - \theta_1| \right| 
\,> \epsilon \right\}$,
the optimal exponent of the second error probability
can be approximately attained.
Note that this measurement depends on the
alternative hypothesis $\rho_{\theta_1}$, and is almost independent of
the null hypothesis $\rho_{\theta_0}$.
This optimality is guaranteed
because the converse part by Ogawa and Nagaoka \cite{ON}
is valid in this case.
Thus, it implies 
\begin{eqnarray*}
B^{\vec{M}_{\theta_1}}
( \rho_{\theta_0} \|\rho_{\theta_1} )
=D( \rho_{\theta_0} \|\rho_{\theta_1} ),
\end{eqnarray*}
where $\vec{M}_{\theta_1}:= \{ M^n_{\theta_1}\}$.
\section{Application of information-spectrum method}\Label{s2}
In order to prove Theorems \ref{thm9} and \ref{thm1},
we have to treat general sequences of probabilities 
having no structure like a Markov chain
because the sequence of two probabilities
${\rm P}_{\rho^{\otimes n}}^{M^n}$ and 
${\rm P}_{\sigma^{\otimes n}}^{M^n}$ generally have no structure.
In the classical information theory,
Han\cite{Han,Han2} introduced the information-spectrum method
in order to treat a general sequence of information sources.
In this section, 
we simply review the information-spectrum method 
in classical hypothesis testing,
and by using this method,
we characterize $B^{\vec{M}} ( \rho \| \sigma)$
and $B^{\dagger,\vec{M}} ( \rho \| \sigma)$.
Given two general sequences of probabilities
$\vec{p}= \{p_n\}$ and
$\vec{q}=\{ q_n\}$
on the same probability sets $\{\Omega_n\}$,
we may define the general hypothesis testing problem 
with $\vec{p}= \{p_n\}$ as the null hypothesis
and $\vec{q}=\{ q_n\}$ as the alternative hypothesis.
In this situation, Any {\it classical test} 
is described by a function $T^n: \Omega_n \to [0,1]$.
For any test $T^n$, the error probabilities of
the first and the second are, respectively, defined by
\begin{eqnarray*}
\alpha_n(T^n):=
\sum_{\omega_n \in \Omega_n}
(1- T^n(\omega_n))p_n(\omega_n) ,
\quad
\beta_n(T^n):=
\sum_{\omega_n \in \Omega_n}
T^n(\omega_n)q_n(\omega_n) .
\end{eqnarray*}
We focus on the two following quantities
\begin{eqnarray*}
\lo B ( \vec{p} \| \vec{q})
:=
\sup \left\{
\lambda \left|
\exists \vec{A},\quad
\lim_{n \to \infty} \alpha_n(T^n) = 0,\quad
 \limsup_{n \to \infty}
\frac{1}{n} \log \beta_n(T^n) \le -\lambda 
\right. \right\} , \\
\lo B^{\dagger} ( \vec{p} \| \vec{q})
:=
\sup \left\{
\lambda \left|
\exists \vec{A},\quad
\liminf_{n \to \infty} \alpha_n(T^n) \,< 1,
\quad
\limsup_{n \to \infty}
\frac{1}{n} \log \beta_n(T^n) \le -\lambda 
\right.\right\} ,
\end{eqnarray*}
which can be regarded as generalizations of
$B^{\vec{M}} ( \rho \| \sigma)$ and $B^{\dagger,\vec{M}} ( \rho \| \sigma)$.
In the independent identical distributed (i.i.d.) case
of $p$ and $q$, as is known as Stein's lemma,
these two values $B ( \vec{p} \| \vec{q})$ and $B^{\dagger}( \vec{p} \| \vec{q})$
coincide with the relative entropy (Kullback-Leibler divergence)
$D(p\|q)$.
Since the relative entropy $D(p\|q)$ is 
the expectation of the variable $\log \frac{p}{q}$ 
under the distribution $p$,
these two values $B ( \vec{p} \| \vec{q})$ and $B^{\dagger}( \vec{p} \| \vec{q})$ 
seem related to the variable 
$\frac{1}{n}\log \frac{p_n}{q_n}$ 
under the distribution $p_n$.
In order to characterize the asymptotic 
behavior of the variable 
$\frac{1}{n}\log \frac{p_n}{q_n}$ 
under the distribution $p_n$,
we define other two values by
\begin{eqnarray*}
\underline{D}(\vec{p}\|\vec{q})&:=
\sup \left\{ \lambda \left|
\lim_{n \to \infty}
p_n \left\{ \omega_n\left| \frac{1}{n} \log \frac{p_n(\omega_n)}{q_n(\omega_n)}
\,< \lambda \right\}\right.
= 0 \right\}\right. , \\
\overline{D}(\vec{p}\|\vec{q})&:=
\inf \left\{ \lambda \left|
\lim_{n \to \infty}
p_n \left\{ \omega_n\left| \frac{1}{n} \log \frac{p_n(\omega_n)}{q_n(\omega_n)}
\,> \lambda \right\}\right.
= 0 \right\}\right. .
\end{eqnarray*}
As guaranteed by the following lemma, 
two values $B( \vec{p}\| \vec{q})$ and $B( \vec{p}\| \vec{q})$ 
are characterized by 
the asymptotic behavior of the variable 
$\frac{1}{n}\log \frac{p_n}{q_n}$ 
under the distribution $p_n$.
\begin{lem}[Han \cite{Han},Verd\'{u} \cite{Verdu},
Nagaoka \cite{Naga,Naga2}]
\Label{L0}
We can show the relations
\begin{eqnarray}
B(\vec{p}\|\vec{q})&= \underline{D}(\vec{p}\|\vec{q})\Label{J1} \\
 B^{\dagger}(\vec{p}\|\vec{q})&=\overline{D}(\vec{p}\|\vec{q}) \Label{J2} .
\end{eqnarray}
Defining a test $T^n(\lambda)$ as the test with the acceptance region
$S_n(\lambda)$:
\begin{eqnarray*}
S_n(\lambda):=
\left\{ \omega_n\left|
\frac{1}{n}\log \frac{p_n(\omega_n)}{q_n(\omega_n)}
\ge \lambda\right\}\right.  ,
\end{eqnarray*}
we obtain 
\begin{eqnarray}
\alpha_n(T^n(\underline{D}(\vec{p}\|\vec{q})-\epsilon)) &\to 0 \Label{201}\\
\beta_n(T^n(\underline{D}(\vec{p}\|\vec{q})-\epsilon))
&\le e^{-n (\underline{D}(\vec{p}\|\vec{q})-\epsilon)} \Label{202}\\
\liminf_{n \to \infty}
\alpha_n(T^n(\overline{D}(\vec{p}\|\vec{q})-\epsilon))
&\,< 1 \Label{203}\\
\beta_n(T^n(\overline{D}(\vec{p}\|\vec{q})-\epsilon))
&\le e^{-n (\overline{D}(\vec{p}\|\vec{q})-\epsilon)} \Label{204}
\end{eqnarray}
for any $\epsilon \,> 0$.
Thus, the tests 
$T^n(\underline{D}(\vec{p}\|\vec{q})-\epsilon)$
and $T^n(\overline{D}(\vec{p}\|\vec{q})-\epsilon)$
approximately attain the bounds
$ \underline{D}(\vec{p}\|\vec{q})$ and 
$ \overline{D}(\vec{p}\|\vec{q})$, respectively.
\end{lem}
The equation (\ref{J1}) was proven in Chapter 4 in Han\cite{Han}.
He referred to Verd\'{u}\cite{Verdu}.
The equation (\ref{J2}) was 
derived by Nagaoka\cite{Naga,Naga2}.
Nagaoka and Hayashi \cite{NH} contains Lemma \ref{L0} and a comprehensive
review of the information-spectrum method.
For reader's convenience, we give a proof in Appendix A.

Next, in order to apply Lemma \ref{L0} to the characterization 
of $B^{\vec{M}} ( \rho \| \sigma)$
and $B^{\dagger,\vec{M}} ( \rho \| \sigma)$,
we define
\begin{eqnarray*}
\underline{D}^{\vec{M}}
(\rho\|\sigma) 
:= 
\underline{D}\left( \left. \left\{ 
{\rm P}^{M^n}_{\rho^{\otimes n}}\right\} \right\|
\left\{ {\rm P}^{M^n}_{\sigma^{\otimes n}}\right\}\right),
\quad
\overline{D}^{\vec{M}}
(\rho\|\sigma) 
:= \overline{D}\left( \left. \left\{ 
{\rm P}^{M^n}_{\rho^{\otimes n}}\right\} \right\|
\left\{ {\rm P}^{M^n}_{\sigma^{\otimes n}}\right\}\right)
\end{eqnarray*}
for any sequence $\vec{M}$ of POVMs.
From Lemma \ref{L0}, we have
\begin{eqnarray*}
\underline{D}^{\vec{M}}(\rho\|\sigma) 
=B^{\vec{M}}(\rho\|\sigma) , \quad
\overline{D}^{\vec{M}}(\rho\|\sigma)
=B^{\dagger,\vec{M}}(\rho\|\sigma) .
\end{eqnarray*}
Therefore, 
a sequence $\vec{M}$ of POVMs
is optimal in the sense of Stein's lemma
if and only if 
\begin{eqnarray*}
\underline{D}^{\vec{M}}(\rho\|\sigma)= D(\rho\|\sigma).
\end{eqnarray*}
In the following, we discuss 
$\underline{D}^{\vec{M}}(\rho\|\sigma)$ and
$\overline{D}^{\vec{M}}(\rho\|\sigma)$
instead of
$B^{\vec{M}}(\rho\|\sigma)$ and 
$B^{\dagger,\vec{M}}(\rho\|\sigma)$.

In this paper, we use only Lemma \ref{L0} among 
several results regarding the information-spectrum method,
and this lemma is sufficient for our current issue.
This paper treats only an application of 
the classical information-spectrum method
to quantum hypothesis testing,
while Nagaoka and Hayashi \cite{NH} discussed 
a quantum analogue of the information-spectrum method.
The references, Han \cite{Han,Han2} and Nagaoka and Hayashi \cite{NH}
may be useful for
the reader who is interested in 
other related topics concerning the
information-spectrum method.

\section{Proof of main result}\Label{sp}
In this section,
by using Lemma \ref{L0},
we prove Theorems \ref{thm9} and \ref{thm1}.
In our proofs, we use 
Lemma \ref{725} and
several fundamental inequalities
given in section \ref{snew}.

\noindent{\it Proof of Theorem \ref{thm9}:}\quad 
As is guaranteed by Lemma \ref{L0},
it is sufficient to show
\begin{eqnarray*}
\underline{D}^{\vec{M}}(\rho\|\sigma)
= D ( \rho \| \sigma ) \Label{721}.
\end{eqnarray*}
First, we prove that 
the variable $\frac{1}{n}\log {\rm P}_{\rho^{\otimes n}}^{M^n}$ 
converges to $\Tr \rho \log \rho$ in probability.
We can calculate
\begin{eqnarray*}
\fl \sum_{i}{\rm P}_{\rho^{\otimes n}}^{M^n}(i)\left(
\frac{1}{n}\log {\rm P}_{\rho^{\otimes n}}^{M^n}(i)
- \Tr \rho \log \rho \right)^2 \\
\fl =
\Tr {\cal E}_{M^n}(\rho^{\otimes n}) \left(
\frac{1}{n}
\log {\cal E}_{M^n}(\rho^{\otimes n}) -  
\Tr \rho \log \rho \right)^2
= \Tr \rho^{\otimes n} \left(
\frac{1}{n}\log {\cal E}_{M^n}(\rho^{\otimes n}) -  
\frac{1}{n}\Tr \rho \log \rho \right)^2\\
\fl \le  2 \Tr \rho ^{\otimes n}
\left(\frac{1}{n}\log {\cal E}_{M^n}(\rho^{\otimes n}) - 
\frac{1}{n} \log \rho^{\otimes n} \right)^2
+ 2 \Tr \rho^{\otimes n}\left(
\frac{1}{n}\log \rho^{\otimes n} -  \Tr \rho \log \rho \right)^2
 \\
\fl \le 8 \left(\frac{ (k-1)\log ( n+1) }{n}\right)^2 
+ 2 \Tr \rho^{\otimes n}
\left(\frac{1}{n}(\log \rho)^{(n)} 
- \Tr \rho \log \rho \right)^2  \\
\fl = 8 \left(\frac{ (k-1)\log ( n+1) }{n}\right)^2 
+ \frac{2}{n}
 \Tr \rho \left( \log \rho
- \Tr \rho \log \rho \right)^2 , 
\end{eqnarray*}
where the last inequality follows from 
Lemma \ref{lem2} given in section \ref{snew}.
Thus, the variable $\frac{1}{n}\log {\rm P}_{\rho^{\otimes n}}^{M^n}$ 
converges to $\Tr \rho \log \rho$ in probability.

Since 
$\frac{1}{n}\log \frac
{{\rm P}_{\rho^{\otimes n}}^{M^n}}{{\rm P}_{\sigma^{\otimes n}}^{M^n}}
=
\frac{1}{n}\log {{\rm P}_{\rho^{\otimes n}}^{M^n}}- 
\frac{1}{n}\log {{\rm P}_{\sigma^{\otimes n}}^{M^n}}
$ and
$D(\rho\|\sigma)= \Tr \rho \log \rho - \Tr \rho \log \sigma$,
the condition (\ref{721}) is equivalent to
\begin{eqnarray}
- \Tr \rho \log \sigma
= \sup\left\{
\lambda \left|
\lim_{n \to \infty} {\rm P}_{\rho^{\otimes n}}^{M^n}\left\{
-\frac{1}{n}\log {\rm P}_{\sigma^{\otimes n}}^{M^n} \,< \lambda \right\}
=0 \right. \right\} \Label{726}.
\end{eqnarray}
Since
\begin{eqnarray*}
\fl
\sup\left\{
\lambda \left|
\lim_{n \to \infty} {\rm P}_{\rho^{\otimes n}}^{M^n}\left\{
-\frac{1}{n}\log {\rm P}_{\sigma^{\otimes n}}^{M^n} \,< \lambda \right\}
=0 \right. \right\} 
\le
\inf\left\{
\lambda \left|
\lim_{n \to \infty} {\rm P}_{\rho^{\otimes n}}^{M^n}\left\{
-\frac{1}{n}\log {\rm P}_{\sigma^{\otimes n}}^{M^n} \,> \lambda \right\}
=0 \right. \right\} ,
\end{eqnarray*}
it follows from Lemma \ref{725} that
the condition (\ref{726}) equivalent to
\begin{eqnarray*}
\fl - \Tr \rho \log \sigma
= \sup\left\{
\lambda \left|
\lim_{n \to \infty} {\rm P}_{\rho^{\otimes n}}^{M^n}\left\{
-\frac{1}{n}\log {\rm P}_{\sigma^{\otimes n}}^{M^n} \,< \lambda \right\}
=0 \right. \right\} \\
\lo =  
\inf\left\{
\lambda \left|
\lim_{n \to \infty} {\rm P}_{\rho^{\otimes n}}^{M^n}\left\{
-\frac{1}{n}\log {\rm P}_{\sigma^{\otimes n}}^{M^n} \,> \lambda \right\}
=0 \right. \right\} .
\end{eqnarray*}
Therefore, if and only if 
the variables $- \frac{1}{n}\log {\rm P}_{\sigma^{\otimes n}}^{M^n}$ 
converges to $- \Tr \rho \log \sigma$ in probability,
the relation (\ref{721}) holds.
\endproof

Next, we give Lemma \ref{725} with a proof,
which is used in our proof of Theorem \ref{thm9}.
\begin{lem}\Label{725}
Under the same assumption as Theorem \ref{thm9},
we obtain 
\begin{eqnarray}
- \Tr \rho \log \sigma
\ge 
\inf\left\{
\lambda \left|
\lim_{n \to \infty} {\rm P}_{\rho^{\otimes n}}^{M^n}\left\{
-\frac{1}{n}\log {\rm P}_{\sigma^{\otimes n}}^{M^n} \,> \lambda \right\}
=0 \right. \right\} . \Label{724}
\end{eqnarray}
\end{lem}
\begin{proof}
We discuss the asymptotic behavior of 
the variable $\frac{1}{n}\log {\rm P}_{\sigma^{\otimes n}}^{M^n}$.
From Markov inequality,
we have 
\begin{eqnarray*}
p \{ X \ge a \} 
\le e^{- \Lambda(X,p,a)} , \quad
\Lambda(X,p,a):= \sup_{0 \le t \le 1} 
\left(a t - \log \int e^{t X(\omega)} p(\,d \omega) \right) .
\end{eqnarray*}
We can calculate 
\begin{eqnarray*}
\fl  \Lambda\left(- \log {\rm P}_{\sigma^{\otimes n}}^{M^n},
{\rm P}_{\rho^{\otimes n}}^{M^n},a n \right) 
= \sup_{0 \le t \le 1} a n t - \log \Tr 
\left({\cal E}_{M^n}(\rho^{\otimes n})
\left({\cal E}_{M^n}(\sigma^{\otimes n})\right)^{-t} \right) \\
\fl = \sup_{0 \le t \le 1} a n t - \log \Tr 
\left(\rho^{\otimes n}
\left({\cal E}_{M^n}(\sigma^{\otimes n})\right)^{-t} \right) \\
\fl \ge   \sup_{0 \le t \le 1} a n t - \left(  t \log w(E^n)  
+ \log \Tr \rho^{\otimes n} \left(\sigma^{\otimes n}\right)^{-t} \right)\\
\fl =  \sup_{0 \le t \le 1}  n \left( a t - t \frac{\log w(E^n)  }{n}
- \log \Tr \rho \sigma^{-t} \right),
\end{eqnarray*}
where the inequality $\ge$ follows from Lemma \ref{c2} given 
in section \ref{snew}.
If $a \,> - \Tr \rho \log \sigma$, then 
there exists a real number $t_0$
such that $0 \,< t_0 \,< 1$
and 
$\frac{\log \Tr \rho \sigma^{-t_0} }{t_0} \,< a
$ because 
$\lim_{t \to 0}
\frac{\log \Tr \rho \sigma^{-t} }{t} = 
- \Tr \rho \log \sigma$.
Therefore the inequalities
\begin{eqnarray*}
\fl \liminf_{n \to \infty}
\frac{-1}{n}\log 
{\rm P}_{\rho^{\otimes n}}^{M^n}\left\{
-\frac{1}{n}\log {\rm P}_{\sigma^{\otimes n}}^{M^n} \,> a \right\} \\
\fl \ge
\liminf_{n \to \infty}
\Lambda\left(- \log {\rm P}_{\sigma^{\otimes n}}^{M^n},
{\rm P}_{\rho^{\otimes n}}^{M^n},a n \right)  
\ge
\lim_{n \to \infty}
\sup_{0 \le t \le 1} \left( a t - t \frac{(k+1)\log (n+1)  }{n}
- \log \Tr \rho \sigma^{-t} \right) \\
\fl = \sup_{0 \le t \le 1} \left( a t - \log \Tr \rho \sigma^{-t} \right) 
\ge t_0( a- \frac{\log \Tr \rho \sigma^{-t_0} }{t_0})
\,> 0
\end{eqnarray*}
hold, i.e., 
\begin{eqnarray*}
\lim 
{\rm P}_{\rho^{\otimes n}}^{M^n}\left\{
-\frac{1}{n}\log {\rm P}_{\sigma^{\otimes n}}^{M^n} \,> a \right\} 
= 0.
\end{eqnarray*}
Thus, inequality (\ref{724}) holds. 
\end{proof}
\noindent{\it Proof of Theorem \ref{thm1}:}\quad 
From equation (\ref{e919}) and Lemma \ref{L0},
it is sufficient to prove 
the inequality
\begin{eqnarray*}
D(\rho\|\sigma) \ge 
\overline{D}^{\vec{M}}(\rho\|\sigma)
\end{eqnarray*}
for any sequence $\vec{M}=\{M^n\}$ such that
$M^n \ge E^n$ and $w(M^n)=1$.
As shown in our proof of Theorem \ref{thm9},
the variable $\frac{1}{n}\log {\rm P}_{\rho^{\otimes n}}^{M^n}$ 
converges to $\Tr \rho \log \rho$ in probability.
Therefore,
\begin{eqnarray*}
\overline{D}^{\vec{M}}(\rho\|\sigma) 
= \Tr \rho \log \rho
+
\inf\left\{
\lambda \left|
\lim_{n \to \infty} {\rm P}_{\rho^{\otimes n}}^{M^n}\left\{
-\frac{1}{n}\log {\rm P}_{\sigma^{\otimes n}}^{M^n} \,> \lambda \right\}
=0 \right. \right\}  \\
\le
\Tr \rho \log \rho -\Tr \rho \log \sigma = D(\rho \|\sigma),
\end{eqnarray*}
where the inequality follows from (\ref{724}).
\endproof

\section{Fundamental inequalities}\Label{snew}
In this section,
we give some fundamental inequalities
used in our proofs of Theorem \ref{thm9} and Lemma \ref{725}.
\begin{lem}\Label{lem2}
If PVMs $E,M$ satisfy $M \ge E$
and a state $\rho$ is commutative with $E$
nd $w(E) \ge 3$, then
the inequality
\begin{eqnarray}
\Tr \rho
(\log \rho - \log {\cal E}_{M}(\rho))^2
\le 4(\log w(E))^2 \Label{62} 
\end{eqnarray}
holds.
\end{lem}
\begin{proof}
Define $a_i :=\Tr E_i \rho E_i ,\rho_i: = \frac{1}{a_i} E_i \rho E_i$,
then the equations
$\rho = \sum_{i}a_i \rho_i,
{\cal E}_M(\rho) =  \sum_{i}a_i{\cal E}_M(\rho_i)$
hold.
Using the operator inequality
$(A+B)^2 \le 2 (A^2 + B^2)$,
we have 
\begin{eqnarray*}
\Tr \rho
(\log \rho - \log {\cal E}_{M}(\rho))^2 
=
\sum_{i} a_i 
\Tr \rho_i
(\log \rho_i - \log {\cal E}_{M}(\rho_i))^2  \\
\lo \le 
\sup_{i} 
\Tr \rho_i
(\log \rho_i - \log {\cal E}_{M}(\rho_i))^2 
\le 
\sup_i
\Tr \rho_i
2 \left( (\log \rho_i)^2 +(\log {\cal E}_{M}(\rho_i))^2 \right) \\
\lo = 
2\sup_i\left(
\Tr \rho_i 
(\log \rho_i)^2 
+\Tr{\cal E}_{M}(\rho_i) (\log {\cal E}_{M}(\rho_i))^2 \right)
\le  4 \sup_i(\log \dim E_i)^2,
\end{eqnarray*}
where the last inequality follows from Lemma \ref{2jou}.
We obtain (\ref{62}).
\end{proof}
\begin{lem}[Nagaoka\cite{Naga3}, Osawa\cite{Osawa}]\Label{2jou}
The equation 
\begin{eqnarray}
\lo  \max \left\{\left.  \sum_{i=1}^k p_i (\log p_i)^2 \right|
 p_i \ge 0, \sum_{i=1}^k p_i= 1\right\} \nonumber \\
\lo = 
\left\{
\begin{array}{cc}
 (\log k)^2 & \hbox{ if } k \ge 3 \\
\frac{1-\sqrt{1-\frac{4}{e^2}}}{2}
\left( \log \frac{1-\sqrt{1-\frac{4}{e^2}}}{2}\right)^2
+
\frac{1+\sqrt{1-\frac{4}{e^2}}}{2}
\left( \log \frac{1+\sqrt{1-\frac{4}{e^2}}}{2}\right)^2
& \hbox{ if } k = 2
\end{array}
\right.
\Label{22}
\end{eqnarray}
holds.
\end{lem}
Its proof is given in Appendix B.
\begin{lem}\Label{c1}
Let $k$ be the dimension of ${\cal H}$.
For any state $\rho \in {\cal S}({\cal H})$
and any PVM $M$,
the inequality 
$\rho \le k {\cal E}_M( \rho ) $ holds.
\end{lem}
\begin{proof}
The relations
\begin{eqnarray*}
\fl \left\langle \psi \left| \left(
{\cal E}_M( | \phi \rangle \langle \phi | ) k 
- | \phi \rangle \langle \phi |  
\right)\right| \psi \right\rangle 
=  k \sum_{i=1}^k
 \langle \psi |M_i | \phi \rangle \langle \phi |M_i  | \psi \rangle
- \left| \sum_{i=1}^k \langle \psi |M_i | \phi \rangle \right|^2 \ge 0 
\end{eqnarray*}
hold for $\forall \phi, \forall \psi \in {\cal H}$,
where the inequality follows from 
Schwartz' inequality about vectors
 $\{ \langle \psi |M_i | \phi \rangle 
\}_{i=1}^k, \{1 \}_{i=1}^k$.
Thus, we obtain
$| \phi \rangle \langle \phi |
\le k {\cal E}_M( | \phi \rangle \langle \phi | )$.
Any state $\rho$ can be decomposed as
$\rho = \sum_i s_i | \phi_i \rangle \langle \phi_i |$.
Thus,
\begin{eqnarray*}
\rho = \sum_i s_i | \phi_i \rangle \langle \phi_i |
\le  \sum_i s_i k {\cal E}_M( | \phi_i \rangle \langle \phi_i | )=
k {\cal E}_M(\rho).
\end{eqnarray*}
The proof is completed.
\end{proof}
\begin{lem}\Label{c2}
Let $\rho$ be a state commuting the PVM $E$.
If the PVM $M$ satisfies $M \ge E$,
the operator inequality 
\begin{eqnarray}
w(E)^{t}\rho^{-t} \ge \left({\cal E}_M( \rho )\right)^{-t} \Label{99}
\end{eqnarray}
holds for $0 \,< t \le 1$ when $\rho^{-1}$ is bounded.
\end{lem}
\begin{proof}
Based on the same notation as in our proof of Lemma \ref{lem2},
it follows from Lemma \ref{c1} that
\begin{eqnarray*}
\rho = \sum_i a_i \rho_i 
\le  \sum_i a_i \rank E_i {\cal E}_M( \rho_i ) 
\le  \sum_i a_i w(E) {\cal E}_M( \rho_i ) 
=w(E){\cal E}_M( \rho ) .
\end{eqnarray*}
Since the map $u \to - u^{-t} ~(0 \,< t \le 1)$ is 
an operator monotone function in $(0,\infty)$\cite{Bhatia},
the operator inequality (\ref{99}) holds.
\end{proof}

\section{Relation between $\rho^{\otimes n},\sigma^{\otimes n}$ 
and group representation}\Label{s5}
In this section, by using the representation theory,
we prove the existence
of a PVM $E^n$ such that
the PVM $E^n$ is commutative
with states $\sigma^{\otimes n}$ and $\rho^{\otimes n}$
and $w(E^n) \le (n+1)^{k-1}$.
In subsection \ref{s31}, for this purpose, 
we consider the relation between 
irreducible representations and PVMs.
In subsection \ref{s32}, 
we discuss $n$-tensor product states
from a group theoretical viewpoint
and prove the desired existence.
\subsection{group representation and its irreducible decomposition}\Label{s31}
Let $V$ be a finite dimensional vector space 
over the complex numbers $\complex$.
A map $\pi$ from a group $G$ to the generalized linear group
of a vector space $V$ is called a {\it representation} on $V$
if the map $\pi$ is homomorphism i.e. 
$\pi( g_1 ) \pi (g_2) = \pi ( g_1 g_2 ), ~ \forall g_1 , g_2 \in G$.
A subspace $W$ of $V$ is called {\it invariant} with respect to 
a representation $\pi$ if the vector $\pi (g) w$ belongs to the
subspace $W$ for any vector $w \in W$ and any element $g \in G$.
A representation $\pi$ is called {\it irreducible} if
there is no proper nonzero invariant subspace of $V$
with respect to $\pi$.
Let $\pi_1$ and $\pi_2$ be representations of a group $G$ on 
$V_1$ and $V_2$, respectively.
The {\it tensored} representation $\pi_1 \otimes \pi_2$ of $G$ on $V_1 \otimes V_2$
is defined as
$(\pi_1 \otimes \pi_2) (g) 
= \pi_1  (g) \otimes \pi_2 (g) $,
and the {\it direct sum}
 representation $\pi_1 \oplus \pi_2$ of $G$ on $V_1 \oplus
 V_2$ is also defined as
$(\pi_1 \oplus \pi_2) (g) 
= \pi_1  (g) \oplus \pi_2 (g) $.

In the following,
we treat a representation $\pi$ of a group $G$
on a finite-dimensional Hilbert space ${\cal H}$;
The following facts is crucial in the later arguments.
There exists an irreducible decomposition ${\cal H}=
{\cal H}_1 \oplus \cdots \oplus {\cal H}_l$
such that the irreducible components
are orthogonal to one another
if for any element $g \in G$ 
there exists an element $g^* \in G$ such that
$\pi(g)^*= \pi (g^*)$ where $\pi(g)^*$ denotes the adjoint of 
the linear map $\pi(g)$.
We can regard the irreducible decomposition ${\cal H}=
{\cal H}_1 \oplus \cdots \oplus {\cal H}_l$
as the PVM
$\{ P_{{\cal H}_i} \}_{i=1}^{l}$, where $P_{{\cal H}_i}$ denotes
the projection to ${\cal H}_i$.
If two representations $\pi_1$ and $\pi_2$ satisfy the preceding condition,
then the tensored representation $\pi_1 \otimes \pi_2$, also, 
satisfies it.
Note that, in general,
an irreducible decomposition of a representation satisfying the
preceding condition is not unique.
In other words, we cannot uniquely define the PVM from such a representation.

\subsection{Relation between the tensored representation and PVMs}\Label{s32}
Let the dimension of the Hilbert space ${\cal H}$ be $k$.
Concerning the natural representation $\pi_{\SL({\cal H})}$ of 
the special linear group $\SL({\cal H})$ on ${\cal H}$,
we consider its $n$-th tensored representation
$\pi_{\SL({\cal H})}^{\otimes n}
:= \underbrace{\pi_{\SL({\cal H})} \otimes \cdots 
\otimes \pi_{\SL({\cal H})}}_n$ on the tensored space 
${\cal H}^{\otimes n}$
\cite{Weyl,GW}.
For any element $g \in \SL({\cal H})$,
the relation $\pi_{\SL({\cal H})}(g)^*=
\pi_{\SL({\cal H})}(g^*)$ holds where the element $g^* \in \SL({\cal
  H})$ denotes the adjoint matrix of the matrix $g$.
Consequently, there exists an irreducible decomposition 
of $\pi_{\SL({\cal H})}^{\otimes n}$ regarded as a PVM
and we denote the set of such PVMs by $Ir^{\otimes n}$.
      
From the Weyl's dimension formula 
((7.1.8) or (7.1.17) in Goodman-Wallach\cite{GW}),
the $n$-th symmetric tensored space is
the maximum-dimensional space in 
the irreducible subspaces with respect to the $n$-th tensored representation 
$\pi_{\SL({\cal H})}^{\otimes n}$.
Its dimension equals the repeated combination $~_{k}H_n$
evaluated by 
$~_kH_{n} =  {n+k-1 \choose k-1} = {n+k-1 \choose n}  
=~_{n+1}H_{k-1}\le (n+1)^{k-1} $.
Thus, any element $ E^n\in Ir^{\otimes n}$ satisfies
$w( E^n ) \le (n+1)^{k-1}$.
\begin{lem}\Label{thm2}
A PVM $E^n \in Ir^{\otimes n}$ 
is commutative with the $n$-th tensored state $\rho^{\otimes n}$
of any state $\rho$ on ${\cal H}$.
\end{lem}
\begin{proof}
If $\det \rho \neq 0$,
then this lemma is trivial from the fact that $\det(\rho)^{-1}
\rho \in \SL({\cal H})$.
If $\det \rho = 0$,
there exists a sequence $\{ \rho_i \}_{i=1}^{\infty}$ 
such that $\det \rho_i \neq 0$ and 
$\rho_i \to \rho$ as $i \to \infty$.
We have 
$\rho_i^{\otimes n} \to \rho^{\otimes n}$ as $i \to \infty$.
Because a PVM $E^n \in Ir^{\otimes n}$ 
is commutative with $\rho_i^{\otimes n}$,
it is, also, commutative with $\rho^{\otimes n}$.
\end{proof}
Therefore, the existence of a desired PVM
is proven.
In particular, in the spin 1/2 system,
$E^n$ corresponds to the measurement of the total momentum.
Therefore, $E^n \times E({\sigma^{\otimes n}})$ 
can be regarded as a simultaneous measurement 
of the total momentum and a momentum of the specified direction.

\section{Conclusion}
We discuss quantum measurements from a viewpoint of
quantum hypothesis testing.
We characterize a sequence of quantum measurements
whose second error exponent attains the quantum relative entropy
at Theorem \ref{thm9} in the finite-dimensional case.
As a byproduct, 
we give another proof of quantum Stein's lemma.
This characterization is closely related to
the irreducible decomposition of the tensored representation
of the group $\SL({\cal H})$.
In our proof of the main theorem, the information-spectrum method 
plays an important role.
In the further research of quantum information,
this method seems a powerful and useful technique.
In addition, as a special case of the infinite-dimensional case,
we treat the quantum Gaussian states.
The photon counting measurement is used in the construction
of our asymptotically optimal measurement,
and this fact indicates its importance.
\section*{Acknowledgment}
The author wishes 
to thank Professor H. Nagaoka, Mr. S. Osawa, Dr. T. Ogawa 
and Dr. Y. Watanabe for useful comments.

\appendix 
\section{Proof of Lemma \ref{L0}}
We simplify $\underline{D}(\vec{p}\|\vec{q})$ and 
$\overline{D}(\vec{p}\|\vec{q})$ by $\underline{D}$ and $\overline{D}$,
respectively.

{\it Direct part $\ge$ of (\ref{J1}):}
For any $\epsilon \,> 0$,
we have
\begin{eqnarray*}
\alpha_n(T^n(\underline{D}-\epsilon))
= p_n (S_n(\underline{D}-\epsilon)^c)
=
p_n 
\left\{ \omega_n\left| \frac{1}{n} \log \frac{p_n(\omega_n)}{q_n(\omega_n)}
\,<\underline{D}-\epsilon
 \right\}\right. \to 0
\end{eqnarray*}
and
\begin{eqnarray*}
\beta_n(T^n(\underline{D}-\epsilon))
&= q_n
\left\{ \omega_n\left| \frac{1}{n} \log \frac{p_n(\omega_n)}{q_n(\omega_n)}
\ge\underline{D}-\epsilon
 \right\}\right.\\
&\le
e^{-n (\underline{D}-\epsilon)} p_n
\left\{ \omega_n\left| \frac{1}{n} \log \frac{p_n(\omega_n)}{q_n(\omega_n)}
\ge\underline{D}-\epsilon
 \right\}\right.
\le e^{-n (\underline{D}-\epsilon)},
\end{eqnarray*}
which imply (\ref{201}) and (\ref{202}).
Thus, 
\begin{eqnarray*}
\limsup_{n \to \infty}
\frac{1}{n}
\log \beta_n(T^n(\underline{D}-\epsilon))
\le
- (\underline{D}-\epsilon).
\end{eqnarray*}

{\it Direct part $\ge$ of (\ref{J2}):}
Note that
\begin{eqnarray*}
\overline{D}=
\sup \left\{ \lambda \left|
\liminf_{n \to \infty}
p_n \left\{ \omega_n\left| \frac{1}{n} \log \frac{p_n(\omega_n)}{q_n(\omega_n)}
\le \lambda \right\}\right.
\,< 1 \right\}\right. .
\end{eqnarray*}
For any $\epsilon \,> 0$,
similarly, 
we have
\begin{eqnarray*}
\liminf_{n \to \infty}
\alpha_n(T^n(\overline{D}-\epsilon))
&=
\liminf_{n \to \infty} p_n 
\left\{ \omega_n\left| \frac{1}{n} \log \frac{p_n(\omega_n)}{q_n(\omega_n)}
\,<\overline{D}-\epsilon
 \right\}\right. \,< 1 \\
\beta_n(T^n(\overline{D}-\epsilon))
&\le e^{-n (\overline{D}-\epsilon)},
\end{eqnarray*}
which imply (\ref{203}) and (\ref{204}).
Thus, 
\begin{eqnarray*}
\liminf_{n \to \infty}
\frac{1}{n}
\log \beta_n(T^n(\overline{D}-\epsilon))
\le
- (\overline{D}-\epsilon).
\end{eqnarray*}

{\it Converse part $\le$ of (\ref{J1}):}
Assume that $\alpha_n(T^n) \to 0 $ as $n \to \infty$
and 
\begin{eqnarray*}
\limsup_{n \to \infty}
\frac{1}{n}\log \beta_n(T^n)= - R.
\end{eqnarray*}
For any $\epsilon \, > 0$,
the inequality
\begin{eqnarray}
\fl \alpha_n(T^n(R-\epsilon))+ e^{n(R-\epsilon)}\beta_n(T^n(R-\epsilon)) 
=
1
+ \sum_{\omega_n} ( 
e^{n(R-\epsilon)}q_n(\omega_n) - p_n(\omega_n) )
T^n(R-\epsilon) (\omega_n) \nonumber \\
\fl \le
1
+ \sum_{\omega_n} ( 
e^{n(R-\epsilon)}q_n(\omega_n) - p_n(\omega_n) )
T^n (\omega_n)
= \alpha_n(T^n) + e^{n(R-\epsilon)}\beta_n(T^n) \Label{NP}
\end{eqnarray}
holds.
Since the RHS goes to $0$ and
$e^{n(R-\epsilon)}\beta_n(T^n(R-\epsilon)) \ge 0$,
the relation \begin{eqnarray*}
p_n \left\{ \omega_n\left| \frac{1}{n} \log \frac{p_n(\omega_n)}{q_n(\omega_n)}
\,<R -\epsilon
 \right\}\right.
= \alpha_n(T^n(R-\epsilon))
\to 0
\end{eqnarray*}
holds. It implies that $R - \epsilon \,< \underline{D}$.

{\it Converse part $\le$ of (\ref{J2}):}
Assume that $\liminf_{n \to \infty}\alpha_n(T^n) \,< 1 $
and 
\begin{eqnarray}
 \limsup_{n \to \infty}
\frac{1}{n}\log \beta_n(T^n)= - R. \Label{J22}
\end{eqnarray}
For any $\epsilon \, > 0$,
from (\ref{NP}) and (\ref{J22}),
we have
\begin{eqnarray*}
\fl \liminf_{n \to \infty}
p_n \left\{ \omega_n\left| \frac{1}{n} \log \frac{p_n(\omega_n)}{q_n(\omega_n)}
\,<R -\epsilon
 \right\}\right.
= \liminf_{n \to \infty}\alpha_n(T^n(R-\epsilon))
\le \liminf_{n \to \infty}\alpha_n(T^n)\,< 1.
\end{eqnarray*}
It implies that
$R - \epsilon \,< \overline{D}$.

\section{Proof of Lemma \ref{2jou}}
In the cases $k=2,3$, the equation
(\ref{22}) is checked by a calculation.
Now, we prove (\ref{22}) by induction
in the case $k \ge 4$.
Let $a_k$ be the RHS of (\ref{22}).
The inequality $a_k \ge (\log k)^2$ is trivial.
From the assumption of the induction,
if $a_k=   \sum_{i=1}^k p_i (\log p_i)^2$,
then $p_i \,> 0~(i=1, \ldots , k)$.
Using Lagrange multiplier method,
we have 
$(\log p_i )^2+ 2 \log p_i - \lambda' =0$,
where $\lambda'$ is the Lagrange multiplier.
The solution is written by 
$\log p_i= -1 \pm \lambda$,
where $\lambda:= \sqrt{1+ \lambda'}$.
Without loss of generality,
we can assume that 
there exists $0\le r \le k$ such that
\begin{eqnarray*}
\log p_i = \left\{
\begin{array}{cc}
-1 + \lambda & \hbox{ if } r \ge i \\
-1 - \lambda & \hbox{ if } r \,< i . 
\end{array}\right.
\end{eqnarray*}
Since $\sum_i p_i =1$,
we have 
\begin{eqnarray*}
1= r e^{-1+\lambda} + (k-r) e^{-1-\lambda},
\end{eqnarray*}
which is equivalent to the 
quadratic equation
\begin{eqnarray*}
rx^2 - e x + k -r =0,
\end{eqnarray*}
where $x:=e^{\lambda}$.
Since the discriminant is greater than $0$,
we have 
\begin{eqnarray*}
e^2 - 4r(k-r) \ge 0,
\end{eqnarray*}
which is solved as:
\begin{eqnarray}
r \le \frac{k- \sqrt{k^2-e^2}}{2},
 \frac{k+ \sqrt{k^2-e^2}}{2} \le r.
\Label{81}
\end{eqnarray}
The function $c(x):=
\frac{x-\sqrt{x^2-e^2}}{2}$
is monotone decreasing in $(e,\infty)$,
and $c(4) \,< 1$.
Thus, the condition (\ref{81})
implies that $r= 0$ or $k$.
Thus, we have $p_i= 1/k$ i.e., (\ref{22}).

\end{document}